\documentclass[a4paper,fleqn,usenatbib]{mnras}

\usepackage{newtxtext,newtxmath}

\usepackage[T1]{fontenc}
\usepackage{ae,aecompl}
\usepackage{booktabs,threeparttable}

\usepackage{graphicx}	
\usepackage{subcaption}	
\usepackage{amsmath}	
\usepackage{amssymb}	

 \title[POLAMI I: Program and calibration]
        {POLAMI: Polarimetric Monitoring of Active Galactic Nuclei at
                 Millimetre Wavelengths\\
        {\LARGE I. The program, calibration, and calibrators data products}}

\author[I. Agudo et al.]{
            Iv\'an Agudo$^{1}$\thanks{E-mail: iagudo@iaa.es (IA)},
            Clemens Thum$^{2}$,
            Sol N. Molina$^{1}$,
            Carolina Casadio$^{3,1}$ ,
             \newauthor
            Helmut Wiesemeyer$^{3}$,
            David Morris$^{4}$,
            Gabriel Paubert$^{2}$,
            Jos\'e L. G\'omez$^{1}$  
            \newauthor
            and Carsten Kramer$^{2}$  
\\
$^{1}$Instituto de Astrof\'{\i}sica de Andaluc\'{\i}a (CSIC),  
                 Apartado 3004, E--18080 Granada, Spain\\
$^{2}$Instituto de Radio Astronom\'ia Milim\'etrica, 
                Avenida Divina Pastora, 7, Local 20, E--18012 Granada, Spain\\
$^{3}$Max--Planck--Institut f\"ur Radioastronomie, 
                Auf dem H\"ugel, 69, D--53121, Bonn, Germany\\
$^{4}$Institut de Radio Astronomie Millim\'etrique, 
                300 Rue de la Piscine, 38406 St. Martin d'H\`eres, France
}

\date{Accepted XXX. Received YYY; in original form ZZZ}

\pubyear{2017}

\begin{document}
\label{firstpage}
\pagerange{\pageref{firstpage}--\pageref{lastpage}}
\maketitle

\begin{abstract}
We describe the POLAMI program for the monitoring of all  four Stokes parameters of a sample of bright radio--loud active galactic nuclei with the IRAM 30\,m telescope at 3.5 and 1.3\,mm. 
The program started in October 2006 and accumulated, until August 2014,  2300 observations at 3.5\,mm, achieving a median time sampling interval of 22 days for the sample of 37 sources.
This first paper explains the source selection, mostly blazars, the observing strategy and data calibration, and gives the details of the instrumental polarisation corrections. 
The sensitivity (1$\sigma$) reached at 3.5\,mm is $0.5$\,\% (linear polarisation degree),  $4.7^{\circ}$ (polarisation angle), and $0.23$\,\% (circular polarisation), while the corresponding values at 1.3\,mm are $1.7$\,\%, $9.9^{\circ}$, and $0.72$\,\%, respectively.
The data quality is demonstrated by the time sequences of our calibrators Mars and Uranus. 
For the quasar 3C\,286, widely used as a linear polarisation calibrator, we give improved estimates of its linear polarisation, and show for the first time occasional detections of its weak circular polarisation, which suggests a small level of variability of the source at millimeter wavelengths.
\end{abstract}

\begin{keywords}
Galaxies: active
   -- galaxies: jets
   -- quasars: general 
   -- BL~Lacertae objects: general
   -- polarisation
   -- surveys
\end{keywords}



\section{Introduction}
\label{Intr}

Radio--loud active galactic nuclei (AGN) are powerful emitters of radiation. 
Their relativistic jets radiate at all spectral bands from the longer radio wavelengths to the very high energy $\gamma$-rays.
This emission shows extremely fast and high-amplitude total flux density and polarisation variability throughout the spectrum \citep[e.g.][]{Marscher:2008p15675,Marscher:2010p11374,Abdo:2010p11811,Agudo:2011p15946,Aleksic:2015p23952,Casadio:2015p27231}.

Even after several decades of study since their discovery there are still relevant questions about some of the most fundamental physics driving these objects, especially in the innermost and most poorly explored regions. 
These questions include: which models of jet formation, acceleration and collimation, compression of the plasma, and magnetic field describe these systems best, what is the mechanism at work for the high energy emission, what is the composition of the plasma forming the relativistic jets, or what is the ultimate origin of the variability in the putative accretion and jet ejection system.
The linear and circular polarization spectra, and the strong polarization variability with doubling time scales from hours to months \citep{Saito:2013p27376,Hayashida:2015p27388,Marscher:2016p27369} usually assumed to be produced by the dynamic evolution of inhomogeneities in the plasma forming the jets, are supposed to be intimately related to the answers to these questions.

Total flux density and polarimetric monitoring can therefore help shedding new light on these problems, especially in the short millimeter spectral band where AGN jets were not intensively explored before, and where they are believed to be optically thin (hence free of opacity effects).
At radio wavelengths, AGN jets also tend to be affected by Faraday rotation.  
For rotation measures { up to $|\rm{RM}|\approx10^4$\,rad\,m$^{-2}$ sometimes} encountered in AGN jets { at centimetre wavelengths} \citep{Zavala:2004p138,Hovatta:2012p18733}, the observed polarization angle $\chi_{\rm{obs}}=\chi_{\rm{int}}+\rm{RM}~\lambda^{2}$  can be rotated away from the intrinsic polarization angle $\chi_{\rm{int}}$ by several tens, even hundreds, of degrees. 
However, such rotation is usually negligible in the short millimetre range.
This also makes millimetre observations less sensitive to Faraday de--polarisation, hence reflecting better the intrinsic linear polarization properties, and therefore those of the magnetic field, of the system.

Previous programs that focused on the millimetre emission of large samples of AGN have been dedicated to studies of the time dependence of the total flux density and linear polarisation on typical time scales much longer than a month. 
Little has been published on the study of circular polarisation, or the evolution of relatively big AGN samples on shorter time scales ($\sim2$\,weeks).
The only previous attempt to study the variability of the full--polarimetric (i.e. 4 Stokes parameters) properties of a large sample of AGN at millimetre wavelengths was presented in \cite{Agudo:2014p22485}, where we investigated the variations of the 3.5\,mm total flux density, linear and circular polarisation parameters of a set of $\sim100$ AGN with the IRAM 30\,m telescope in 2010 and in 2005 \citep[see][]{Agudo:2010p12104,Agudo:2014p22485}.
The results from this variability study showed prominent total flux density and polarisation changes in time scales of a few years.
Such changes reached median factors of $\sim$1.5 in total flux density and $\sim$1.7 in linear polarisation degree (with maximum variations by factors up to 6.3 and $\sim$5, respectively). 
We also noted drastic variations of the polarisation angle, with random changes in 86\% of the sources with regard to the first survey in 2005.
	
In light of these large changes in most sources between our two surveys we embarked on a higher cadence polarisation monitoring program of a selected group of AGN. 
The program, dubbed POLAMI (Polarimetric Monitoring of AGN at Millimetre Wavelengths, see \href{url}{http://polami.iaa.es}), started in 2006, and it measures all four Stokes parameters observed with the IRAM 30\,m Telescope at both 3.5 and 1.3\,mm wavelengths.
In this first paper of a series presenting the results of the first 8 years of POLAMI observations, we provide detailed information about the observing program (Section~\ref{Prog}), the monitored source sample (Section~\ref{Samp}), the data reduction and calibration (Section~\ref{s:Red}), and we demonstrate the quality of our data by showing the results obtained for our main calibrators (Section~\ref{Res}).
The data obtained from our science targets, as well as an analysis of their circular polarisation properties and of their total flux density and linear polarisation variability are shown in the accompanying Paper II 
\citep{PaperII}
and Paper III 
\citep{PaperIII}, respectively. 
Detailed studies of specific sources and discussions of statistical aspects of our sample, like correlations with optical, $\gamma$--ray, and millimetre VLBI data will be presented in separate publications, currently in preparation.

\section{The program}
\label{Prog}

The POLAMI program for the study of the polarimetric properties of AGN at short millimetre wavelengths evolved from technical work which monitored the calibration of the flux density scale and the polarisation characteristics of the IRAM 30\,m Telescope. 
Observations of bright and point--like sources were recorded with complete polarisation information, and it quickly became  clear that this data set constituted a unique resource for AGN studies at short mm wavelengths.

The variability of AGN across the electromagnetic spectrum requires studies that monitor large samples over long time baselines. 
Unprecedented time sampling of the $\gamma$--ray and X--ray emission became available from space observatories, as well as exhaustive optical polarimetric monitoring and millimetre VLBI monitoring of large sets of AGN. 
The combination of such comprehensive multi--spectral--range data sets has been crucial in understanding the innermost, and most poorly explored regions of jets in AGN, where the most energetic processes take place, \citep[e.g.][]{Marscher:2010p11374,Jorstad:2010p11830,Jorstad:2013p21321,Agudo:2011p14707,Agudo:2011p15946, Casadio:2015p27231}.
The POLAMI program fits very well into this broad--band effort. 
In fact, it bridges a gap at short millimetre wavelengths where no systematic full--polarization AGN monitoring programs exist. 

The first POLAMI observing session took place on 14--Oct--2006.
In this first publication we present the data compiled until 18--Aug--2014, but the program is still active and more recent data will be analyzed on already planned publications.
Observations were scheduled in 4--6-hour sessions scattered very irregularly over this $\sim8$ year period. 
The median time sampling of our program $t_\mathrm{S}$ is 22 days, when all monitored AGN targets and all observations are considered over the monitoring period. 
$t_\mathrm{S}$ improved with time, dropping from $\sim42$\,days during the first 4 years to $\sim19$\,days towards the end. 
Observations were interrupted during a brief period in 2009  from February to mid June when a new generation of receivers (Section~\ref{ss:Obs}) was installed and commissioned.

All POLAMI observations are made at sky frequencies of 86.243\,GHz (3.5\,mm) and at 228.932\,GHz (1.3\,mm). 
At both frequencies the performance of the receivers is good (see section~\ref{ss:Obs}), and the atmosphere is free of major absorption lines.
Data are recorded throughout our program using the XPOL procedure \citep{2008PASP..120..777T}. 
All 4 Stokes parameters are obtained strictly simultaneously. 
In December 2009, the 3.5 and 1.3\,mm bands (hereafter 3 and 1\,mm, respectively) started to be observed in parallel, and XPOL then simultaneously recorded all Stokes parameters at both frequencies.

In a typical observing session, 15--20 sources are observed. 
Each polarimetric measurement of a target source is preceded by a pointing observation made on the same source.
Integration times of the polarimetric measurement range from  4 min (sources of 10 Jy or brighter) to 15 min duration for the weaker sources in our sample (see sect.~\ref{Samp}) with $\sim1$ Jy. 
Whenever possible, observations of calibrators were included in each session. 
Mars and Uranus are used for instrumental polarization and total flux density calibration, while 3C\,286 and the Crab nebula allowed for cross--checks of the polarisation calibration.  
Instrumental polarisation parameters  and total power calibration were sufficiently stable to allow interpolation for the few sessions without a calibrator measurement (see section~\ref{s:Red}).

Until mid August 2014, we accumulated 2700 polarimetric observations at 3\,mm. 
After necessary data quality checks, we retain 2300 observations.  
All data, including pointing scans preceding every polarization integration, are compiled in a dedicated database\footnote{The POLAMI database is installed at IRAM--Granada's computer system and is maintained by one of the authors (CT) supported by the Observatory's computer group.}. 
For each observation, the database stores the values of the 4 Stokes parameters and their uncertainties, information about observing conditions, and a quality flag, for each frequency band. 
Polarisation data are kept both in their original form before rotation to the celestial reference frame and before any instrumental polarisation is removed, as well as in their final form. 
Successive refinements of data analysis are then feasible without loss of information. 

\section{The source sample}
\label{Samp}

The entire POLAMI source sample is listed in Table~\ref{sample} together with the most relevant properties of every source.
The sample is designed as the intersection of two different larger samples. 
The first of them, i.e. the IRAM 30\,m Telescope's Pointing Source Catalog \citep{Agudo:2010p12104}, is a set of bright millimetre AGN with declination $\gtrsim-25^{\circ}$, which are easily visible to the 30\,m Telescope.
The second one, is the source sample of the VLBA--BU--BLAZAR Monitoring Program\footnote{\tt http://www.bu.edu/blazars/VLBAproject.html} \citep{Jorstad:2016p25971,Jorstad2017inpress} which compiles sequences of monthly 7\,mm polarimetric VLBI images since 2007.
The intersection of the two sets includes 35 of the 37 sources currently monitored with VLBI by the VLBA--BU--BLAZAR Program, which happen to be among the brightest $\gamma$-ray AGN in the northern sky, and are therefore also intensively monitored by the \emph{Fermi} Gamma Ray Space Observatory, the Swift X-ray satellite, and a number of optical \citep{Agudo:2012p17697,Angelakis:2016p27451,Larionov:2016p25966,Itoh:2016p26051} and radio \citep{Aller:2003p4761,Lister:2009p5316,Terasranta:2004p6726,MaxMoerbeck:2010p26215} total flux density and polarimetric monitoring programs.
This helps maximizing the scientific output of the POLAMI program.
To these 35 sources, we added two more bright millimetre AGN, namely {NRAO\,150} and {3C\,286}, that we considered of interest to our astrophysical and astronomical objectives.
Although {NRAO\,150} is close to the Galactic plane, and therefore its high Galactic absorption prevented its identification and classification until recently \citep{AcostaPulido:2010p13756}, this source is one of the brightest blazars in the northern sky \citep{Agudo:2007p132} and shows interesting polarimetric behaviour on its sub-milliarcsecond structure as revealed by 7\,mm and 3\,mm VLBI monitoring observations \citep{Agudo:2007p132,Molina:2014p22482}.
The quasar {3C\,286} is a standard total flux density and linear polarisation calibrator both at centimetre \citep{Baars:1977p16842,Ott:1994p16851,Perley:1982p6054} and millimetre wavelengths \citep{Agudo:2012p17464}, and has therefore been an excellent control source for our program.
        
All observations were subject { of} a detailed quality control as described in section~\ref{ss:bad}. The median number of accepted valid measurements per source at 3\,mm is 61.
The source with most measurements (108) is the BL\,Lac object 0716+714, and the source with the least number of observations (25) is 1101+384.
The POLAMI source sample comprises 24 radio quasars, 11 BL~Lacertae type objects\footnote{BL~Lac objects hereafter.}, and 3 radio galaxies.
Among the 11 BL~Lac objects, most of them (7) are low--energy synchrotron peaked (LSP), 2 of them are intermediate--energy synchrotron peaked (ISP), and the 2 remaining ones have their synchrotron bump peaking at high energies (HSP).
The redshift of the sources in the POLAMI sample ranges from $z=0.0176$ (for {3C\,84}) to $z=2.218$ (for {4C~71.07}), and has a median of $z=0.815$.

\begin{table*}
\caption{List of sources in our POLAMI first priority sample and their most relevant properties.}
\label{sample}      
\centering  
\begin{tabular}{ccccccccccc}
\hline             
Source & Common & Position ref. & Optical & SED   &  Redshift & \multicolumn{2}{c}{\,$\langle S \rangle$ [Jy]} & \multicolumn{2}{c}{\,N$_{\rm obs}$}\\   
 name  & name  &   & class   & class &          & {\small 3\,mm}  &   {\small 1\,mm}    &  \small 3\,mm   &   \small 1\,mm      \\    
(1) &  (2) &  (3) &  (4) &  (5) &  (6) &  (7) &  (8)   & (9) & (10) \\
\hline 
0219+428  &        3C\,66A & $^{a}$ &  B                   &  HSP &  0.444  &   0.54 &   0.49 &   31  &   7  \\
0235+164  &   AO\,0235+164 & $^{a}$ &  B                   &  LSP &  0.94   &   1.61 &   0.91 &   50  &  15  \\
0316+413  &         3C\,84 & $^{b}$ &  G                   &  LSP &  0.0176 &  18.19 &   8.88 &   59  &  29  \\
0336-019  &        CTA\,26 & $^{a}$ &  Q                   &  LSP &  0.852  &   2.25 &   1.31 &   37  &  19  \\  
0355+508  &      NRAO\,150 & $^{b}$ &  Q$^{e}$  &  LSP &  1.517$^{e}$       &   4.62 &   1.91 &   76  &  35  \\
0415+379  &        3C\,111 & $^{d}$ &  G                   &  LSP &  0.0491 &   3.70 &   2.44 &   91  &  47  \\
0420-014  &   PKS\,0420-01 & $^{a}$ &  Q                   &  LSP &  0.9161 &   4.21 &   2.28 &   46  &  29  \\
0430+052  &        3C\,120 & $^{a}$ &  G                   &  LSP &  0.033  &   2.21 &   1.37 &   26  &  14  \\ 
0528+134  &  PKS\,0528+134 & $^{a}$ &  Q                   &  LSP &  2.07   &   2.76 &   1.60 &   68  &  26  \\
0716+714  &    S5\,0716+71 & $^{a}$ &  B                   &  ISP &  0.127  &   3.78 &   2.87 &  108  &  64  \\
0735+178  &        OI\,158 & $^{a}$ &  B                   &  LSP &  0.45   &   0.87 &   0.61 &   52  &  31  \\
0827+243  &        OJ\,248 & $^{a}$ &  Q                   &  LSP &  0.942  &   1.52 &   1.93 &   73  &  42  \\
0829+046  &        OJ\,049 & $^{a}$ &  B                   &  LSP &  0.174  &   0.70 &   0.53 &   52  &  30  \\
0836+710  &      4C\,71.07 & $^{a}$ &  Q                   &  LSP &  2.218  &   1.48 &   0.65 &   91  &  48  \\
0851+202  &        OJ\,287 & $^{a}$ &  B                   &  LSP &  0.306  &   5.52 &   3.63 &   92  &  57  \\
0954+658  &    S4\,0954+65 & $^{a}$ &  B        &  LSP & $\geq$0.45$^{f}$   &   1.36 &   0.97 &   79  &  42  \\
1055+018  &      4C +01.28 & $^{a}$ &  Q                   &  LSP &  0.888  &   4.10 &   2.30 &   61  &  38  \\
1101+384  &       Mrk\,421 & $^{d}$ &  B                   &  HSP &  0.0308 &   0.50 &   0.32 &   25  &  18  \\
1127-145  &   PKS\,1127-14 & $^{a}$ &  Q                   &  LSP &  1.187  &   1.64 &   0.72 &   56  &  28  \\
1156+295  &      4C\,29.45 & $^{a}$ &  Q                   &  LSP &  0.725  &   1.37 &   0.82 &   59  &  32  \\
1219+285  &       W\,Comae & $^{a}$ &  B                   &  ISP &  0.103  &   0.44 &   0.34 &   33  &  17  \\
1222+216  &     4C\,+21.35 & $^{b}$ &  Q                   &  LSP &  0.434  &   1.65 &   1.01 &   60  &  45  \\
1226+023  &        3C\,273 & $^{a}$ &  Q                   &  LSP &  0.1583 &  11.61 &   4.08 &   83  &  53  \\
1253-055  &        3C\,279 & $^{a}$ &  Q                   &  LSP &  0.536  &  20.71 &  12.35 &   77  &  37  \\
1308+326  &        OP\,313 & $^{a}$ &  Q                   &  LSP &  0.997  &   1.66 &   0.97 &   54  &  35  \\
1328+307  &        3C\,286 & $^{c}$ &  Q                   &  LSP &  0.846  &   0.90 &   0.31 &   57  &  26  \\
1406-076  &  PKS\,1406-076 & $^{a}$ &  Q                   &  LSP &  1.494  &   0.75 &   0.36 &   32  &  14  \\   
1510-089  &   PKS\,1510-08 & $^{a}$ &  Q                   &  LSP &  0.36   &   2.75 &   1.89 &   53  &  27  \\
1611+343  &        DA\,406 & $^{a}$ &  Q                   &  LSP &  1.401  &   2.18 &   1.13 &   69  &  43  \\
1633+382  &      4C\,38.41 & $^{a}$ &  Q                   &  LSP &  1.813  &   4.11 &   2.85 &   76  &  48  \\
1641+399  &        3C\,345 & $^{a}$ &  Q                   &  LSP &  0.593  &   3.65 &   1.76 &   80  &  52  \\
1730-130  &      NRAO\,530 & $^{a}$ &  Q                   &  LSP &  0.902  &   3.13 &   1.50 &   48  &  26  \\
1749+096  &     4C\,+09.57 & $^{a}$ &  B                   &  LSP &  0.322  &   3.22 &   1.70 &   62  &  44  \\
2200+420  &   BL\,Lacertae & $^{a}$ &  B                   &  LSP &  0.0686 &   6.98 &   5.59 &   70  &  44  \\
2223-052  &        3C\,446 & $^{a}$ &  Q                   &  LSP &  1.404  &   4.58 &   1.21 &   60  &  29  \\
2230+114  &       CTA\,102 & $^{a}$ &  Q                   &  LSP &  1.037  &   3.43 &   1.34 &   67  &  30  \\
2251+158  &      3C\,454.3 & $^{a}$ &  Q                   &  LSP &  0.859  &  24.20 &  24.22 &   80  &  39  \\
\hline                                        
\end{tabular} 
\newline
\begin{tablenotes}
\item Columns are as follows:
(1) {IAU B1950 source name},
(2) {common (alternative) source name},
(3) {reference from where position was taken (see below references $^{a}$ to $^{d}$},
(4) {optical classification into quasars (Q), BL\,Lac objects (B), and radio galaxies (G) after \citet{VeronCetty:2006p4900}},
(5) {classification of sources based on the broad-band spectral-energy distribution (low, intermediate, and, high synchrotron peaked are LSP, ISP, and HSP, respectively) following \citet{Ackermann:2011p17255,Ackermann:2015p22989} and the information provided in the MOJAVE database (\href{url}{http://www.physics.purdue.edu/astro/MOJAVE/})},
(6) {redshift from the MOJAVE database},
(7) {average total flux density measured at 3\,mm},
(8) {average total flux density measured at 1\,mm},
(9)  {number of valid observations at 3\,mm},
(10) {number of valid observations at 1\,mm}.
The number of 1\,mm measurements is significantly smaller than the one at 3\,mm. This is mainly because of the { shorter} time span of 1\,mm monitoring observations (that started in December 2009, see text), but also because of the larger fraction of rejected data at 1\,mm. At this wavelength, atmospheric instabilities are more pronounced, and sources are often considerably weaker than at 3\,mm.
\item $^{a}${\citet{Johnston:1995p27787}}
\item $^{b}${\citet{Beasley:2002p27806}}
\item $^{c}${\citet{Ma:1998p27818}}
\item $^{d}${\citet{Fey:2004p27857}}
\item $^{e}${From \citet{Agudo:2007p132}  and \citet{AcostaPulido:2010p13756}.}
\item $^{f}${From \cite{Landoni:2015p24553}.}
\end{tablenotes}
\end{table*}

\section{Observation and calibration}
\label{s:Red}

\subsection{Observation and antenna temperature calibration}
\label{ss:Obs}

The XPOL observing procedure \citep{2008PASP..120..777T} makes use of the Observatory's horizontally (H) and vertically (V) polarised heterodyne receivers whose down-converted signals are detected in the backend VESPA after auto--correlations (Stokes $I$ and $Q$)  and cross--correlation ($U$ and $V$).
Stokes $I$ and $Q$ are calibrated like any spectroscopic observation using hot and cold loads. 
The system temperature $T_{sys}$ of the $U$ and $V$ spectra are derived from those of the H and V spectra as $T_{sys} = \sqrt{T_{sys}^H \cdot T_{sys}^V}$ \citep{1991isra.book.....T}. 

Each polarimetric observation of a source, typically of 3 -- 15 min duration, was preceded by a pointing observation on the same source. 
Parameters of the pointing were recorded for an assessment of the quality of the polarimetric observation, that was based on the stability of the cross--scan profiles during the time spanned by the pointing measurements.
The pointing accuracy of the telescope, $<1''$ for our observing strategy\footnote{\scriptsize {See \href{url}{http://www.iram.es/IRAMES/telescope/telescopeSummary/telescope\_summary.html}}}, is negligible with regard to other effects affecting the quality of the observations, e.g. the atmospheric stability (see below).
    
During the course of our monitoring program, in spring 2009, the receivers labelled ABCD (after the 4 different cryostats in which they were 
housed\footnote{\href{url}{http://www.iram.es/IRAMES/mainWiki/AbcdforAstronomers}}
) were replaced by a new receiver EMIR which consists of a single cryostat housing dual polarisation (H and V) receivers for the four atmospheric windows at 3, 2, 1.3, and 0.9mm \citep{2012A&A...538A..89C}.
The warm optics of EMIR allowed simultaneous observations of the 3 and 1.3\,mm bands, an option we used throughout our monitoring project as of December 2009. 
Observing frequencies were 86.2 GHz (AB and EMIR) and 228.9 GHz (EMIR). 
The observing bandwidth, limited by the backend VESPA, was a nominal 640 MHz (AB and EMIR 3\,mm) and 320 MHz (EMIR 1\,mm). 
Receiver temperatures were in the range of 40 -- 60 K (3\,mm) and 80 -- 100 K (1\,mm). 
System temperatures ranged from 100 to 130\,K at 3\,mm and from 150 to 400\,K at 1\,mm.  
Beam widths are $28"$ (3\,mm) and $12"$ (1.3\,mm). 
All types of observing conditions were encountered during the observations which were scheduled as a backup program most of the time. 
The quality of the atmosphere was recorded in the observer's logbook, but also evaluated from the pointing data (Section~\ref{ss:bad}).

\subsection{Total flux density calibration}
\label{ss:TP}

After correcting for the dependence on elevation of the on--axis gain \citep{Greve:1998p24214}, we convert the antenna temperatures (corrected for atmospheric attenuation) to flux densities by applying a Kelvin--to--Jansky conversion factor $\rm{f_{Jy/K}}$. 
We derive $\rm{f_{Jy/K}}$ at the two observing frequencies from observations of {Mars} and {Uranus} taken during the monitoring period and by comparing them with models of the emission spectrum of these two planets. 
The Mars model (Lellouch \& Amri 2006 \footnote{http://www.lesia.obspm.fr/perso/emmanuel-lellouch/mars/}) includes the disk--resolved thermal structure of the planet's surface layers which result in diurnal oscillations of its emitted flux density. 

The brightness temperature of Uranus is strongly wavelength dependent \citep{1993Icar..105..537G}. 
We applied the ESA4 model as used for SPIRE/Herschel photometry \citep{Bendo:2013p26530}\footnote{{\href{url}{ftp://ftp.sciops.esa.int/pub/hsc-calibration/PlanetaryModels/ESA4/}}}, 135.0K (86GHz) and 91.6K (229GHz).
The long term evolution of the planet's 90\,GHz brightness was investigated by \citet{2008A&A...482..359K} who found a seasonal modulation with an amplitude $\leq10$\%. 
The resulting brightness temperature extrapolated to our monitoring period is compatible with the brightness temperature adopted above. 

\begin{figure*}
   \centering
   \includegraphics[width=\textwidth]{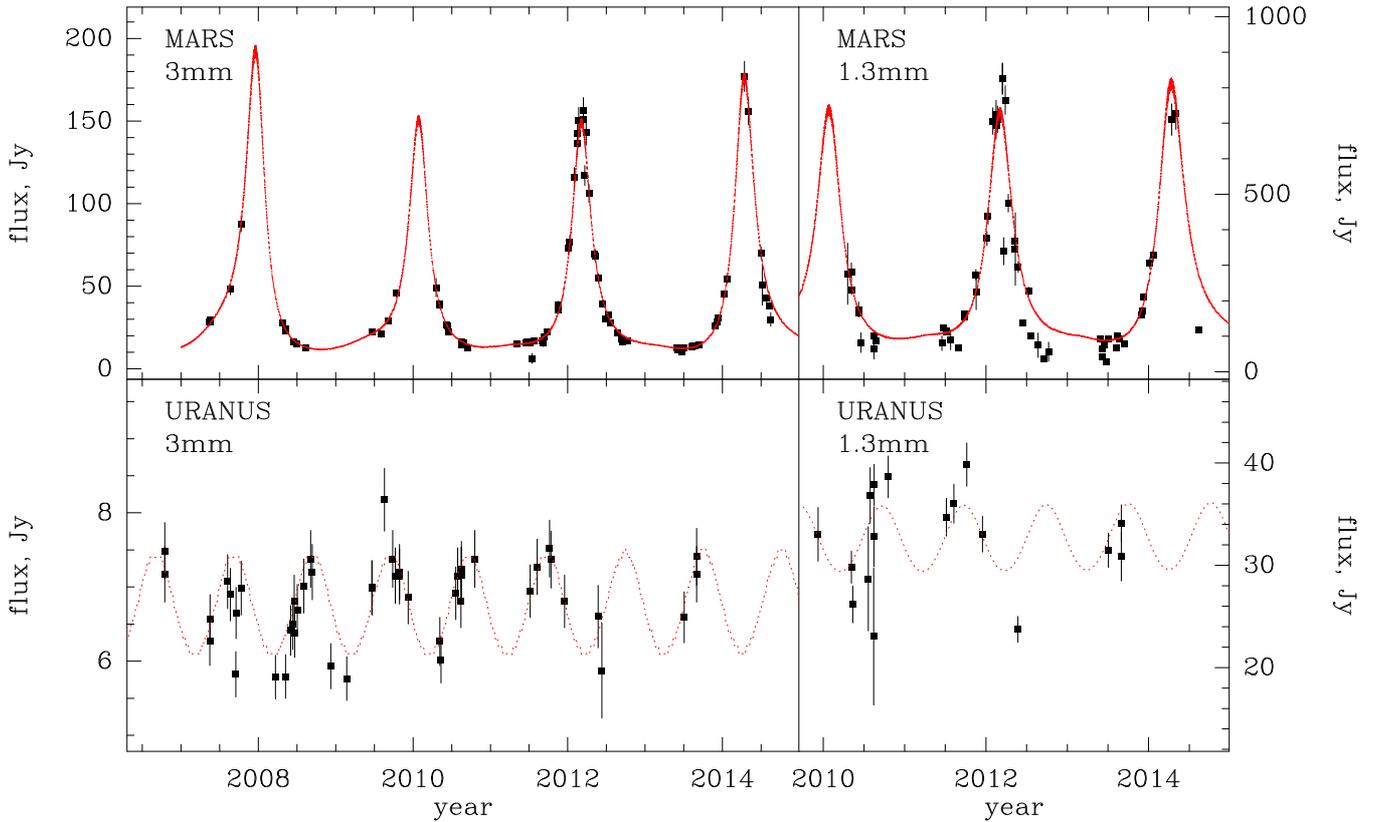}
   \caption{Light curves of Mars (top) and Uranus (bottom) at 3\,mm (left) and 1\,mm (right). Red curves are light curves as predicted by the models after convolution with the planet's disk (see text).}
   \label{f:JyKfact}
\end{figure*}

Figure~\ref{f:JyKfact} shows the flux densities observed at the two wavelengths and the predictions by the models after convolution with the telescope beam\footnote
{For Uranus we used ASTRO which is part of the GILDAS software package. See http://www.iram.fr/IRAMFR/GILDAS}. 
The final $\rm{f_{Jy/K}}$ factors that we applied to calibrate the total flux density, obtained from the comparison of the measurements in Fig.~\ref{f:JyKfact} and the models (see below), are as follows:
\begin{center}
\begin{tabular}{rc}
     $\rm{f_{Jy/K}(3\,mm)}$: &  ($6.1\pm0.1$)\,Jy/K \\
     $\rm{f_{Jy/K}(1\,mm)}$: &  ($8.6\pm0.4$)\,Jy/K  
\end{tabular}
\end{center}
At 3mm wavelength, the results from the two planets agree within measurement error which in turn is comparable to the uncertainties estimated for the models ($\lesssim4$\,\%). The fact that virtually all 3mm observations during the 8 years of the monitoring period are well fitted by a single $\rm{f_{Jy/K}(3\,mm)}$ factor demonstrates the stability of the telescope and its calibration system. 
The increase of Uranus's brightness temperature predicted by the seasonal effect \citep{2008A&A...482..359K} during the monitoring period is very small ($\lesssim2$\,\%) and therefore not seen in our data.

At 1mm, the telescope beam is often comparable or even smaller than the disk of Mars. 
The model prediction then depends on the exact beam shape used, including its near sidelobes, and is therefore less accurate than of the nearly point--like Uranus. 
For our estimate of  $\rm{f_{Jy/K}(1\,mm)}$ we therefore adopt the value derived from Uranus.  

We note that the $\rm{f_{Jy/K}}$ factors derived here are slightly larger than those listed on the IRAM web page\footnote{
http://www.iram.es/IRAMES/mainWiki/Iram30mEfficiencies} which were derived under best observing conditions and refer to an optimum state of the telescope.  
Our data taken under typical observing conditions depart by less than 10\,\% from the optimum values.

The formal errors on the determination of the $\rm{f_{Jy/K}}$ factors do not include the uncertainties in the models, for which we consider a conservative estimate of 5\,\%.
The final uncertainty associated with each total flux density measurement ($\sigma_{S}$, see Table~\ref{t:polErrors}) include a quadratic sum of this 5\,\% (that typically dominates the 3 and 1\,mm error), the error in $\rm{f_{Jy/K}}$, and the statistical uncertainties of the measurement.

\subsection{Instrumental polarisation calibration}
\label{ss:polcal}

\begin{table}
\caption{Evolution of the instrumental polarization parameters of the XPOL polarimeter at the IRAM 30m\, Telescope.}
\label{IPtable}      
\centering  
\begin{tabular}{rcccc}
\hline             
\multicolumn{1}{c}{Time period} &  $Q_{i}$ &  $U_{i}$  &  $V_{i}$ &receiver \\   
                                &   [\%]   &   [\%]    &   [\%]    \\    
\hline
\multicolumn{5}{c}{3\,mm} \\
\hline 
                    2006-10-14  &   -1.13    &   -0.22    &     -0.11 &ABCD \\
                    2006-10-15  &   -4.28    &   -0.04    &      0.10 &ABCD \\
 2007-05-09 $\to$   2007-05-17  &   -2.79    &    0.46    &     -0.15 &ABCD \\
 2007-07-16 $\to$   2007-07-17  &   -1.79    &    0.15    &      0.01 &ABCD \\
 2007-08-07 $\to$   2007-10-11  &   -1.13    &   -0.22    &     -0.11 &ABCD \\
 2008-03-21 $\to$   2008-05-07  &   -0.83    &   -0.18    &     -0.05 &ABCD \\
                    2008-06-01  &   -1.70    &    0.11    &      0.12 &ABCD \\
 2008-06-11 $\to$   2008-06-30  &   -2.34    &   -0.16    &      0.14 &ABCD \\
 2008-07-02 $\to$   2008-09-02  &   -2.89    &   -0.09    &      0.07 &ABCD \\
 2008-09-10 $\to$   2008-11-20  &    0.73    &   -0.18    &      0.18 &ABCD \\
 2008-12-07 $\to$   2008-12-30  &   -1.60    &    0.05    &      0.44 &ABCD \\
 2009-02-16 $\to$   2009-02-20  &   -0.04    &   -0.04    &      0.01 &ABCD \\
                    2009-06-19  &   -2.68    &   -0.17    &  	-0.44 &EMIR \\
 2009-06-20 $\to$   2014-08-18  &   -0.70    &   -0.30    &      0.00 &EMIR \\
\hline                                        
\multicolumn{5}{c}{1\,mm}  \\
\hline 
2009-12-07  $\to$   2011-10-26  &   -0.36    &   -1.07    &      2.57 &EMIR \\
2011-11-16  $\to$   2013-01-15  &   -0.36    &   -1.07    &     -0.42 &EMIR \\
2013-01-29  $\to$   2014-08-18  &   -0.36    &   -1.07    &     -0.92 &EMIR \\
\hline
\end{tabular} 
\end{table}

During the first phase of the monitoring program, when the ABCD receivers were used, the main source of instrumental polarisation (IP) was the small ($\leq2"$), but unstable, misalignment between the horizontal and vertical receivers which were housed in separate dewars. 
Frequent calibration runs (Tab.~\ref{IPtable}) allowed to track the evolution of the IP parameters $Q_i, U_i, {\rm and\ } V_i$ measured in the Nasmyth coordinate system see \citep[see][]{2008PASP..120..777T} where the receivers are located.
Typically, 50 or more observations of unpolarised sources (mainly Mars and Uranus) and the Crab Nebula \citep[the source of well known polarisation properties,][]{Aumont:2010p12769,Wiesemeyer:2011p26387} were made during the time range of our monitoring program. 
After spring 2009, when EMIR was installed, the alignment between orthogonally polarised receivers became stable, since all receivers and their polarisation splitting grids are housed inside a single dewar inaccessible to the outside. 
IP calibration runs had then to be made only after a technical warm up of the dewar.

\begin{figure}
   \centering
   \includegraphics[width=0.8\columnwidth]{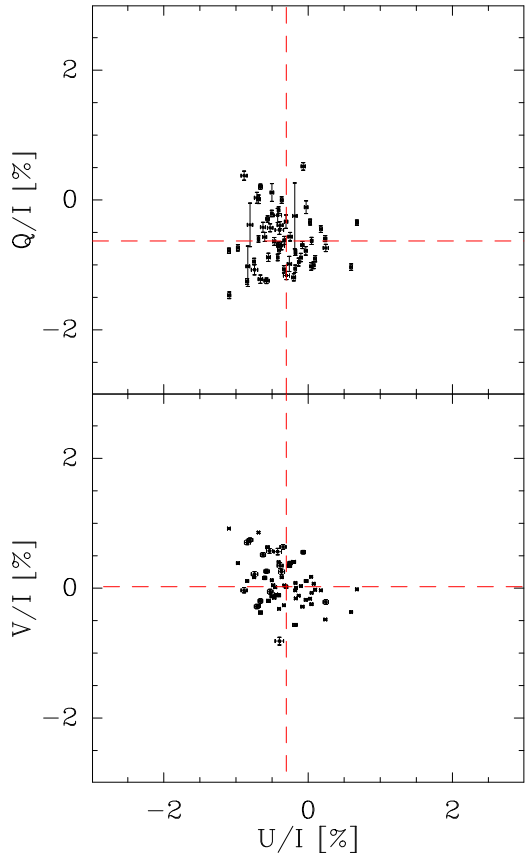}
   \caption{Instrumental polarisation measured at 3\,mm during the time period between 20--Jun--2009 to 18--Aug--2014 from 55 independent measurements of Mars and Uranus. Red dashed lines indicate weighted means of the measurements.}
   \label{IPfig}
\end{figure}

In the data reduction, the IP parameters ($Q_i, U_i, {\rm and\ } V_i$) were subtracted from the Stokes parameters obtained in the Nasmyth system for a given session. 
For sessions in between IP calibration runs,  $Q_i, U_i, {\rm and\ } V_i$ were interpolated. 
The resulting first stage IP--corrected calibrators were then collected in larger time bins with the aim to detect and correct for residual IP errors unnoticed in single calibration runs. 
Table~\ref{t:sysErrors} lists the first stage IP corrections adopted for the entire data set.
Second stage IP corrections were then typically of the order of $\sim0.5$\,\% and  $\sim1$\,\% for 3 and 1\,mm, respectively, and were then subtracted from the target observations.
Figure~\ref{IPfig} shows the results of a second stage IP measurement for the period of 2009 to 2014.   

\begin{table}
\caption{Systematic errors of Stokes parameters (measured from Mars and Uranus as standard variations in the Nasmyth system) due
         to uncontrolled uncertainties in instrumental polarization.}
\label{t:sysErrors}
\begin{center}
\begin{tabular}{lcc}
\hline
 & 3\,mm &1\,mm \\[0.3ex]
 \hline
$\sigma_{Q_{i}}$  & 0.5\,\%  &   1.1\,\% \\
$\sigma_{U_{i}}$  & 0.4\,\%  &   1.7\,\% \\
$\sigma_{V_{i}}$  & 0.2\,\%  &   0.4\,\% \\[0.3ex]
\hline
\end{tabular}
\end{center}
\end{table}

Another potential cause for instrumental polarisation, only affecting Stokes $U$ and $V$ is instrumental phase calibration errors. 
As outlined in \citet{2008PASP..120..777T}, the drift of the phase when using ABCD receivers is not larger than $1^\circ$ per hour which can be easily calibrated.
With EMIR, the phase drift is negligible, making $V$ the potentially best determined Stokes parameter.

We see from  Fig.~\ref{IPfig} that the rms scatter of $Q_i, U_i, {\rm and\ } V_i$ is much larger than the individual errors of the calibrator measurements.
The additional  scatter results in a minor part from residual drifts of the IP, mainly due to uncontrolled small variations in the telescope optics and atmospheric fluctuations, both in brightness and in refractive index. 
Such imperfections, which could well be controlled or eliminated in specific short observing runs, are unavoidable in long term monitoring programs.
We therefore take the rms scatter of the instrumental polarisation measurements of our calibrators (shown in Table~\ref{t:sysErrors}) as a conservative estimate of the uncertainties of instrumental polarisation affecting our data. 
These values were added quadratically to the statistical errors of the Stokes parameters of every single measurement. 
The final error in almost all 3\,mm observations is dominated by these systematic errors.
At 1\,mm, where the system temperature is much higher, the total errors are also dominated by the systematic error, except for sources weaker than about 2\,Jy.

The last step of the polarization calibration is the rotation of the Stokes parameters $Q_N$ and $U_N$ as measured in the Nasmyth system to $Q_{EQ}$ and $U_{EQ}$ in the equatorial system. 
The former was stable enough during the course of our program to guarantee that no additional calibration of the absolute linear polarization angle was needed.
The two sets of Stokes parameters and their respective errors are related by a rotation matrix whose angle $2\tau$ is given in \citep{2008PASP..120..777T}. 
We then derive the typical uncertainties of the linear polarisation degree ($m_{\rm{L}}$), polarisation angle ($\chi$), and circular polarisation  ($m_{\rm{C}}$), estimated as medians over all measurements on our entire sample. 
These uncertainties are given in Table~\ref{t:polErrors}.

 \begin{table}
\caption{{Median errors of the total flux density and polarization data measured in the Equatorial system from the entire source sample defined in Table~\ref{sample}.}}
\label{t:polErrors}
\begin{center}
\begin{tabular}{lcc}
\hline
 & 3\,mm & 1\,mm \\[0.3ex]
 \hline
$\tilde\sigma_{S}$                 & $5.2\,\%$  &   $5.4\,\%$ \\
$\tilde\sigma_{m_{\rm{L}}}$  & $0.51\,\%$  &   $1.71\,\%$ \\
$\tilde\sigma_{\chi}$             & $4.7^{\circ}$  &   $9.9^{\circ}$ \\
$\tilde\sigma_{m_{\rm{C}}}$  & $0.23\,\%$  &   $0.72\,\%$ \\[0.3ex]
\hline
\end{tabular}
\end{center}
\end{table}

\subsection{Data quality control}
\label{ss:bad}

An assessment of the quality of individual observations is based on the observer's logbook and an automated check which evaluates the quality of the pointing scan preceding each polarimetric observation. 
Four criteria were applied to every 4--cross--scan pointing observation: 
{\it (1)} the scatter of the positions derived from Gaussian fits, 
{\it (2)} the scatter of the width of the Gaussians with regard to the nominal beam width at our observing frequencies, 
{\it (3)} the receiver temperature as a diagnosis tool for technical problems, and 
{\it (4)} the system temperature for adverse weather conditions. 
Quality thresholds were set for each of the four criteria, and an observation was rejected whenever one of the criteria was passing its threshold. 
Additionally, if a 3mm observation was rejected, the simultaneous 1mm observation was also rejected.

A particular difficulty arose during periods of increased atmospheric instability when anomalous refraction occurred \citep{1987A&A...184..381A}.
Under these conditions, the telescope beam can be displaced or distorted in ways which are not always caught by our criteria 1 and 2. 
{ Anomalous} refraction events which preferentially occur during summer afternoons and affect observations more at 1mm than at 3mm explain most of the 1mm measurements in Fig.~\ref{f:JyKfact} which fall below the model curves. 
The most severe departure occurs for Mars around 2012.6 where 5 consecutive observations are affected. 
Inspection of these sessions shows however that the { values of} $Q, U$ and $V$ { of the main unpolarized calibrators (Mars and Uranus)} are still acceptable, { i.e. consistent with zero within the errors (see next section)}.
{ The data from these session} albeit with somewhat { larger} errors, were consequently retained. 
Fortunately, the presence of strong anomalous refraction is easily recognised by an experienced observer on bright sources. 
Corresponding entries in the logbooks were used for rejecting individual observations and very occasionally whole observing sessions.

\section{Results}
\label{Res}

\subsection{Unpolarised calibrators}

\begin{figure}
   \centering
   \includegraphics[width=\columnwidth]{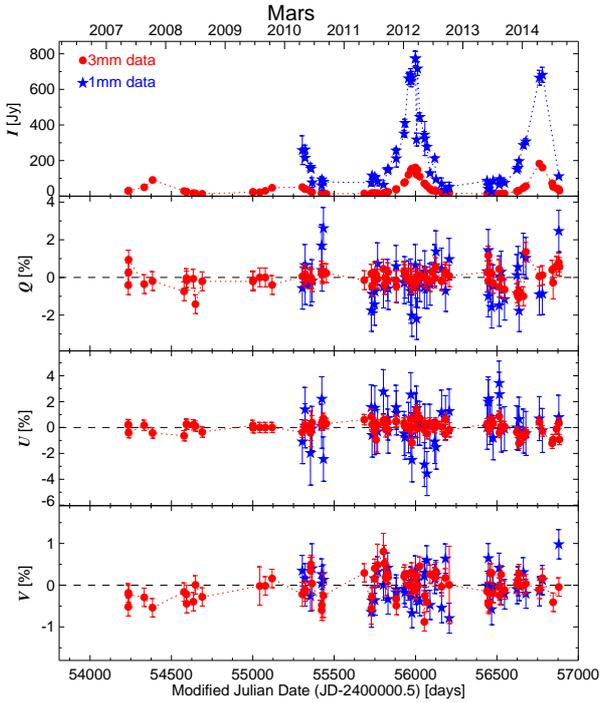}
   \caption{Daily averages of total flux density, and Stokes $Q$, $U$, and $V$ (Nasmyth coordinates) fully calibrated measurements of Mars made by our program at 3 and 1\,mm as a function of time. The clustering of $Q$, $U$, and $V$ measurements of this unpolarised source around 0.0\,\% demonstrates the goodness of our polarisation calibration. The dispersion on the $Q$, $U$, and $V$ measurements reflect the accuracy on the instrumental polarisation determination discussed in Section~\ref{ss:polcal}. The dashed lines on the last three plots show the $Q=0$, $U=0$, and $V=0$ lines, respectively.}
\label{mars}
\end{figure}

\begin{figure}
   \centering
   \includegraphics[width=\columnwidth]{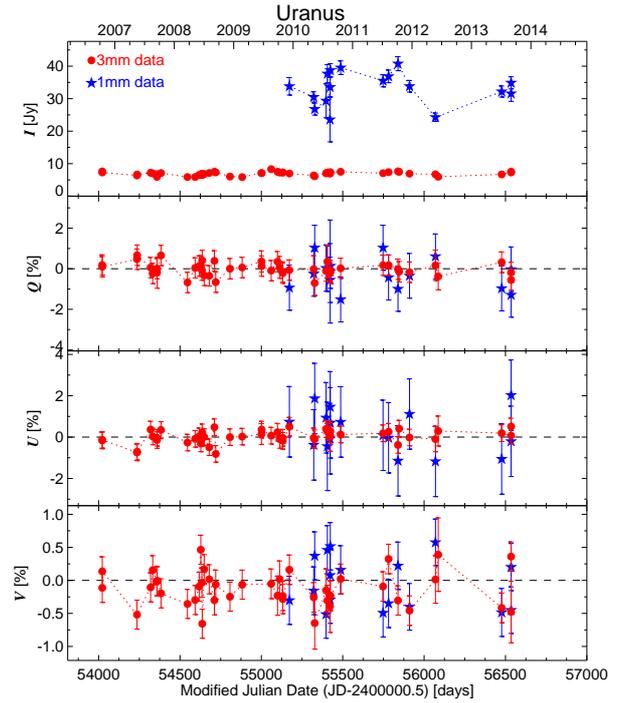}
   \caption{Same as Fig.~\ref{mars} but for Uranus.}
   \label{uranus}
\end{figure}

Figures \ref{mars} and \ref{uranus} show plots of fully calibrated $I$, $Q$, $U$, and $V$ measurements of the Stokes parameters of Mars and Uranus versus time, both for 3 and 1\,mm.
While the Stokes $I$ shows considerable time dependent variability as reproduced by the models shown in Fig.~\ref{f:JyKfact}, the remaining Stokes parameters representing the linear and circular polarisation remain rather constant and consistent with zero.
This is the expected behaviour for any unpolarised source like Mars and Uranus for the IRAM 30\,m Telescope, and constitutes a measure of the goodness of the polarisation calibration, within the uncertainties discussed in Section~\ref{ss:polcal}.

\subsection{Polarised calibrators}

\begin{figure}
   \centering
   \includegraphics[width=\columnwidth]{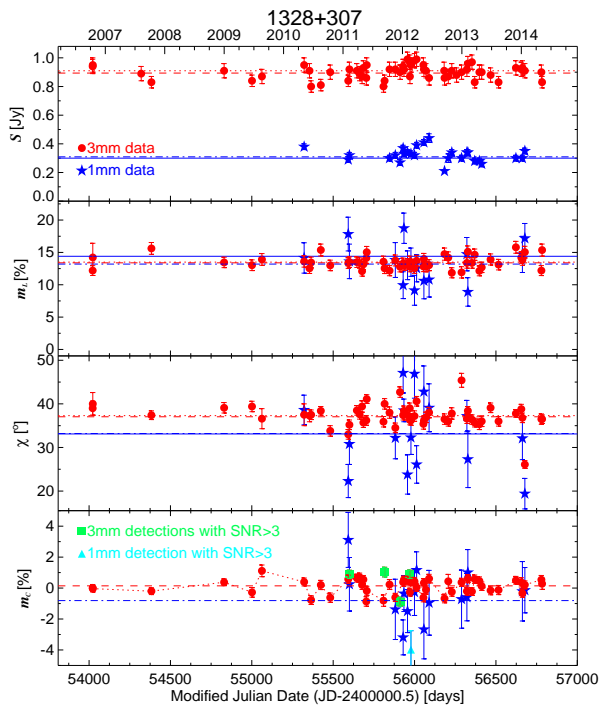}
   \caption{Daily averages of total flux density ($S$), linear polarisation degree ($m_L$), linear polarisation angle ($\chi$), and circular polarisation ($m_C$) of fully calibrated measurements of 3C~286 (1328+307) made by our program as a function of time. The error bars of the 1\,mm total flux density measurements are typically smaller than the size of the corresponding symbols. The dashed lines indicate weighted averages of every 3\,mm data train, whereas dot--dashed lines show the 1\,mm weighted averages. Dotted lines symbolize the 3\,mm weighted averages measured from a shorter data train, as given in \citet{Agudo:2012p17464}, and the continuous lines indicate 1\,mm weighted averaged { data} reported in \citet{Agudo:2012p17464}. Squares and the triangle symbols on the bottom plot show 3\,mm and 1\,mm $m_{\rm{C}}$ detections, respectively, with signal to noise ratio larger than 3.}
   \label{3c286}
\end{figure}

Because of the known total flux density and polarisation stability of 3C~286 at centimetre wavelengths, also demonstrated in the short millimetre regime \citep[][and this paper]{Agudo:2012p17464}, the target was used as a control source for our calibration.
Figure~\ref{3c286} shows the time series of 3 and 1\,mm total flux density ($S$), linear polarisation degree ($m_{\rm{L}}$), linear polarisation angle ($\chi$), and circular polarisation ($m_{\rm{C}}$) of fully calibrated measurements of 3C~286 taken during the course of our program (see also Table~\ref{t:3c286} where we give their weighted averages and standard deviations).
These measurements confirm the results shown in \citet{Agudo:2012p17464} where we demonstrate the low level of variability of the source in $S$, $m_L$, and $\chi$ at both 3 and 1\,mm, and therefore its goodness as total flux density and polarisation calibrator also at short millimetre wavelengths.

 \begin{table}
\caption{Weighted averages and standard deviations of 3 and 1\,mm measurements of $S$, $m_{\rm{L}}$, ${\chi}$, and  $m_{\rm{C}}$ from 3C\,286.}
\label{t:3c286}
\begin{center}
\begin{tabular}{lcc}
\hline
 & 3\,mm & 1\,mm \\[0.3ex]
 \hline
$\langle {S} \rangle$            &$(0.90\pm0.01)$\,Jy & $(0.31\pm0.01)$\,Jy \\
$\sigma_{S}$                   &$0.05$\,Jy&$0.06$\,Jy\\
$\langle {m_{\rm{L}}} \rangle$   &$(13.4\pm0.1)$\,\%&$(13.2\pm0.7)$\,\%\\
$\sigma_{m_{\rm{L}}}$          &$1.0$\,\%&$2.9$\,\%\\
$\langle {\chi} \rangle$         &$(37.0\pm0.4)^{\circ}$&$(33.2\pm2.0)^{\circ}$\\
$\sigma_{\chi}$                &$2.7^{\circ}$&$8.0^{\circ}$\\
$\langle {m_{\rm{C}}} \rangle$   &$(0.14\pm0.07)$\,\%&$(-0.81\pm0.44)$\,\%\\
$\sigma_{m_{\rm{C}}}$          &$0.5$\,\%&$1.8$\,\%\\[0.3ex]
\hline
\end{tabular}
\end{center}
\end{table}

Closer inspection of the circular polarisation data at 3 and 1\,mm, presented here for 3C~286, shows that the observed scatter is nearly twice as high as the median values over the entire sample as given at the end of Section~\ref{ss:polcal}. 
Four 3\,mm observations and one at 1mm have signal--to--noise ratios $\geq3$ and therefore formally qualify as actual $m_{\rm{C}}$ detections. 
Contrary to the other Stokes parameters, circular polarisation in this source does not appear to be stable. 
In Paper II \citep{PaperII}
it is argued that $V$ is the Stokes parameter with the highest variability, and 3C\,286 may be an extreme example.

We also made measurements of the Crab Nebula using the standard observing setup of our program as a check on the overall health of the acquisition system, in particular the sign of Stokes $U$ and the calibration of the phase. 
These measurements successfully reproduced the known linear and circular polarisation properties of the source \citep{Aumont:2010p12769,Wiesemeyer:2011p26387}.

\section{Summary}

In this first of a series of papers, we present the basis of the POLAMI program.
This is an ongoing full-polarisation simultaneous 3 and 1\,mm monitoring program of a set of 37 bright AGN, mostly blazars.
The observations were performed, from October 2006 to August 2014, with the XPOL polarimeter on the IRAM 30\,m Telescope with a median time sampling of 22 days.
We presented here the motivation for this observing program, its strategy, the source sample, and the details regarding the calibration of the data.
In particular, we discuss the total flux density, and instrumental polarisation calibration of the IRAM 30\,m Telescope.
We demonstrate the suitability of the IRAM 30\,m Telescope for full--polarisation observations of blazars at short millimetre wavelengths.   
The polarisation sensitivity is sufficiently high and stable for the successful development of long--term monitoring projects like POLAMI.
We confirm the suitability of 3C286 as a linear polarisation (degree and angle) calibrator, and we present evidence that the source is variable in circular polarisation. 

\section*{Acknowledgements}
     Emmanuel Lellouch (Observatoire de Paris, France) is kindly acknowledged for technical support on the Mars model and its calculations for the time spanned by the POLAMI observations.
     We gratefully acknowledge Emmanouil Angelakis (MPIfR, Germany) for his careful revision and useful comments to improve this manuscript.
     We appreciate the support by the operator team of the IRAM 30\,m telescope. 
     This paper is based on observations carried out with the IRAM 30\,m Telescope and analyzed with GILDAS software.
     IRAM is supported by INSU/CNRS (France), MPG (Germany) and IGN (Spain).
     IA acknowledges support by a Ram\'on y Cajal grant of the Ministerio de Econom\'ia, Industria y Competitividad (MINECO) of Spain.
     The research at the IAA-CSIC was supported in part by the MINECO through grants AYA2016-80889-P, AYA2013-40825-P and AYA2010-14844, and by the regional government of Andaluc\'{i}a through grant P09-FQM-4784.

\bsp	
\label{lastpage}
\end{document}